**Title:** An experimental test for genetic constraints in *Drosophila melanogaster*.

**Authors:** Ian Dworkin[1,2,3], David Tack[1], Jarrod Hadfield[4]

Author affiliations:

    1- Department of Zoology, Michigan State University

    2- Program in Ecology, Evolutionary Biology and Behavior

    3- BEACON Center for the Study of Evolution in Action

    4- Edward Grey Institute of Field Ornithology, University of Oxford

**Corresponding author:** idworkin@msu.edu

## Abstract

The short-term response to selection is determined by the genetic (co)variances between traits – **G**. The long-term response however depends on how **G** evolves. The degree to which **G** reflects patterns of mutational effects (**M**) remains a key question. Yet, **M** is difficult to measure. However, the form of **M** is likely a reflection of the ways in which developmental systems can be perturbed. Thus predicting **G** from knowledge of development offers evidence that a relationship between **M** and **G** remains. Here we introduce a well-characterized mutation into *Drosophila melanogaster* that provides specific predictions about the change in **G** *if* genetic effects can be predicted from known developmental perturbations.

The mutation *Ultrabithorax*[1] causes a homeotic transformation of segmental identity. We predicted that *Ubx*[1] would induce a segment specific change in **G** due to this homeosis. We measured **G** for a set of traits in a panel of strains with and without *Ubx*[1]. As expected, *Ubx*[1] induced homeotic transformations, and altered patterns of allometry yet surprisingly, little change in **G** was observed. We discuss the role of using genetic manipulations to refine hypotheses of constraints in natural systems.

# Introduction

The covariances between phenotype and breeding value encapsulated in the **G**-matrix, determine the direction and rate of evolutionary change in response to selection. Both the relative orientation of **G** with respect to selection ($\beta$), and the dimensionality of **G** itself influence the rate of evolutionary response (Mezey and Houle 2005; McGuigan and Blows 2007; Walsh and Blows 2009). In turn the structure of **G** is influenced by selection, drift and mutation. The structure of **G** for a set of traits is determined by the distribution of mutational effects on those traits and the way in which selection and drift change the frequency of mutants with different pleiotropic effects. Under neutrality, **G** is expected to be proportional to **M**, the mutational covariance matrix, but obtaining theoretical predictions for how **G** depends on selection has been more challenging (Barton and Turelli 1989). Indeed, under persistent directional selection, theoretical models have predicted **G** to be dominated by selection (Charlesworth et al. 1990), whereas others have predicted **G** to be dominated by **M** (Hill 1982; Houle 1991; Barton and de Vladar 2009). Lande (1980) derived equations for the within-generation change in **G** due to selection:

$$\Delta \mathbf{G}^* = \mathbf{G}(\gamma - \beta\beta^T)\mathbf{G}^T$$

where $\beta$ is the vector of directional selection gradients and $\gamma$ the matrix of quadratic selection gradients. Combinations of traits under the strongest directional selection and/or stabilizing selection are expected to exhibit the lowest amount of segregating genetic variance after selection. It is generally argued that in the medium term **G** will be a compromise between **M** and the effects of selection **W** ($-(\gamma - \beta\beta^T)^{-1}$), although the robustness of these conclusions are far from clear given that the extent to which within-generation change in **G** maps to between generation change in **G** is unknown, and that allele frequency changes may be substantial. Indeed, under persistent directional selection, some theoretical models have predicted **G** to be independent of selection (Hill 1982; Houle 1991; Barton and de Vladar 2009) and some simulation studies have found that combinations of traits under the strongest directional selection actually have greater levels of segregating genetic variance due to selective sweeps (Jones et al. 2012). In the absence of directional selection however, the intuitive idea that **G** will lie between **M** and **W** has wider theoretical support (Lande 1980; Cheverud 1984) and has been verified using simulations (Jones et al. 2004).

A great deal of empirical work has gone into testing whether relationships between **G** and patterns of selection exist. Tests looking at the effects of directional selection on the diagonals of **G** have been equivocal with some studies supporting the expected negative relationship (Teplitsky et al. 2009) and others finding the opposite relationship (Blows et al. 2004). Considering **G** in its entirety Blows (2004) found that the direction of directional (sexual) selection was associated with a dimension of **G** in which there was little variance, but failed to find the expected relationship between the matrix of quadratic selection gradients and **G**. Hunt (2007) did however find such a relationship.

In contrast, much less empirical work has gone into understanding the relationship between **M** and **G**, primarily because of the great difficulty in getting precise estimates of **M** (Estes et al. 2005). Although over very long time-scales **M** itself is predicted to evolve towards **W** (Lande 1980, Cheverud 1984), the slow rate at which this happens means that the structure of **M** is likely to reflect developmental pathways more than patterns of selection do. Consequently, observing structure in **G** that reflects these developmental pathways could give us key insights into evolutionary constraints and phenotypic integration imposed by patterns of mutation. Although general theoretical work on the relationships between **G** and developmental rules does exist (Rice 2000; Agrawal et al. 2001; Wolf et al. 2001; Rice 2004), much has focused on resource acquisition-allocation rules (Rendel 1963; Riska et al. 1989; Houle 1991; Worley et al. 2003). Likewise, empirical work has often focused on explaining patterns of genetic co-variances in terms resource acquisition-allocation rules but only rarely have physical, physiological or genetic manipulations being used to test whether this interpretation is justified (e.g. (Rendel 1963; Nijhout and Emlen 1998). Here we use a genetic manipulation to test whether a well characterised homeotic perturbation results in a change in **G** that is consistent with a change in development, thereby testing the feasibility of predicting mutational effects from development.

Mutations in homeotic genes have been observed to transform the fates of one segment into another (Lewis 1978; Morata 1993; McGinnis 1994). In *Drosophila*, mutations in the *Ultrabithorax* (*Ubx*) gene produce a range of homeotic phenotypes (Lewis 1978; Morata and Kerridge 1980; Bender et al. 1983; Sanchez-Herrero et al. 1985; Casanova et al. 1987; Stern 1998; Rozowski and Akam 2002; Stern 2003; Davis et al. 2007), caused by a transformation of the third thoracic segment towards second segmental identities. As an example, this manifests as a transformation of the halteres towards wing-like fates, with severe manifestations including four-winged flies. While this profound developmental change is the result of multiple mutations in the *Ubx* gene, far subtler transformations can also be observed with the *Ubx*[1] mutation, where heterozygotes display less extreme, quantitative transformations such as an increase in haltere size and the presence of occasional bristles similar to those borne by the wing (Gibson and van Helden 1997; Gibson et al. 1999), changes in organ size (Rozowski and Akam 2002; Stern 2003; Davis et al. 2007), and the rare ectopic expression of the apical bristle on the third thoracic leg, where it is not usually present (Rozowski and Akam 2002). Modulation of Ubx function has been implicated in several evolutionary changes between closely related species, as well as potentially at larger taxonomic scales (Stern 1998; Mahfooz et al. 2007; Khila et al. 2009). Previous work has demonstrated that introgression of the *Ubx*[1] mutation into a panel of wild-type lines of *D. melanogaster* reveals considerable segregating variation for modifiers of aspects of the homeotic phenotype (Gibson and van Helden 1997; Gibson et al. 1999), including allelic variation segregating at the *Ubx* locus (Gibson and Hogness 1996; Phinchongsakuldit 2003).

We argue that the weak homeotic transformation produced by *Ubx*[1] provides a useful test for an altered structure in **G**, with clear *a priori* expectations. Given the partial transformation of third thoracic identity towards that of the second due to *Ubx*[1] (Gibson and van Helden 1997), we hypothesize that the legs on T2 (meso-) and T3 (meta-thoracic segment) should be more similar (higher phenotypic and genetic correlations). We demonstrate that *Ubx$^1$* clearly has an effect on the legs of *Drosophila*, and we observe segregating variation for aspects of the homeotic transformation. Yet we observe at best weak evidence for a change in the structure of **G**. We discuss these results within the context of predicting genetic effects from knowledge about development, and suggest future approaches to addressing related questions.

## Material and Methods

**Fly strains and Introgression:** The *Ubx$^1$* allele was obtained from the Bloomington Drosophila stock center. The strains used in this study were as previously described by Dworkin (Dworkin 2005a, b). We backcrossed the *Ubx$^1$* allele into 30 wild iso-female strains using a previously described crossing scheme for other mutations (Dworkin 2005a). For *Ubx$^1$* a consensus phenotype based upon both haltere size and the presence of one or more "wing"-like bristles on the haltere were used to find virgin females heterozygous for the *Ubx$^1$* allele. These females were then used for the proceeding generation of backcrosses to the wild-type strain. Introgressions were performed for 12 generations. After completion of the introgression procedure 26 isofemale strains remained with the *Ubx$^1$* mutation, each paired with a conspecific line without the mutation.

**Experimental rearing:** Five pairs of adult flies (five *Ubx$^1$*/ *Ubx$^+$* females and 5 wild-type *Ubx$^+$*/ *Ubx$^+$* males) from each introgression line were placed into each of two replicate bottles, with standard medium and seeded with live yeast. The flies were allowed to lay eggs for 3 days, and were then transferred to fresh bottles. One set of replicates (two bottles/strain) were reared at 25°C, and the second set incubated at 18°C. Sterilized cotton was placed in each of bottle after 3 days of egg laying providing additional space for pupation. Larval densities were low to moderate for all lines (<100 individuals/bottle). After eclosion, adult flies (*Ubx$^1$*/ *Ubx$^+$* and *Ubx$^+$*/ *Ubx$^+$*) were stored in 70% ethanol. For one isofemale line used (*w*), no flies reared at 18°C eclosed; therefore we examined models with and without this line, with minimal influences on estimates (not shown).

**Dissection and imaging:** All three right legs and the wing were dissected from 10 individuals from each sex/line/temperature/replicate combination. All organs were mounted (70% Glycerol in phosphate buffered saline, with a small amount of phenol added as a preservative), with organs from 10 individuals to a slide. Legs and wings were imaged using an Olympus DP30BW camera mounted on an Olympus BX51 microscope at 40X magnification using Olympus DP controller image capture software (v3.1.1). ImageJ, v1.38 was utilized to obtain morphometric measurements ("measure straight line command"), for three segments of each leg (femur, tibia and basi-tarsus) in addition to the length and width (anterior crossvein) of the wing. For the current study analyses of measurements from the wing were excluded.

## *Analysis*

Unless otherwise described, all analyses were done using R (v.2.13.0), using the libraries lme4 (Bates et al. 2011) and ASREML-R v3.0.1 (Butler 2009).

**The influence of *Ubx$^1$* on the T3 ectopic apical bristle**:

To determine the extent of the homeotic transformation caused by the *Ubx$^1$* mutation, we utilized the presence of the apical bristle on the meta-thoracic leg as a proxy for penetrance of the mutation. We analyzed the data using a logistic regression mixed model with linear predictor:

$$\eta_{ijkl} = \beta_1 + \beta_2 \delta_l + u_k^{(k1)} + \delta_l u_k^{(k2)} + u_{jk}^{(j1)} + \delta_l u_{jk}^{(j2)}$$

for fly i from replicate j of line k in temperature treatment l. $\delta_l$ =1 when the temperature is 25°C and zero when the temperature is 18°C. The β's are fixed effects, the u's and e's random effects. The random effects are superscripted with the term they are associated with (k=line, j=replicate) and the temperature level (1=18°C, 2=25°C). Line effects are assumed to come from a bivariate normal distribution with a mean of zero and estimated covariance matrix. For example, for the k$^{th}$ line effects:

$$\begin{bmatrix} u_k^{(k1)} \\ u_k^{(k2)} \end{bmatrix} \sim N\left(\begin{bmatrix} 0 \\ 0 \end{bmatrix}, \begin{bmatrix} \sigma_{k1}^2 & \sigma_{k1,k2} \\ \sigma_{k1,k2} & \sigma_{k2}^2 \end{bmatrix}\right)$$

Replicate effects were modeled similarly, although the covariance is non-identifiable and were set to zero. Models in which single variances were estimated for each random term (ignoring temperature) or the random term omitted, were also fitted using `lmer` (Laplace approximation) and compared using AIC.

### The influence of *Ubx[1]* on the leg segments of T3 in relation to the T2 wild-type

To address the central question of this study, namely how the homeotic mutation altered the lengths of T3 leg segments and their patterns of covariation we used a multivariate linear mixed model. To achieve this the three leg segments (femur, tarsus and tibia) for each leg-type (T2(Wt), T3(Wt) and T3(*Ubx[1]*)) were treated as separate response variables.

Mean Model:

As fixed terms we had thoracic segment (T2 and T3), leg segment (femur, tarsus and tibia), genotype (Wt and *Ubx[1]*), temperature (18°C and 25°C) and their interactions. The significance of fixed effect terms were assessed using conditional Wald tests, and the significance of single terms was assessed using a two-tailed t-test on the Z-ratio. In both cases the number of lines (26) was used a conservative estimate for the (denominator) degrees of freedom. All effects remained on the same side of the 0.05 significance threshold when the degrees of freedom were set to 2020 (the number of individuals). As described below we additionally verified the results using parametric bootstraps.

Genetic Model:

Given the limited number of genetic lines used for this study we chose not to fit an unstructured genetic (line) covariance matrix for the nine traits (leg segment by leg-type). Instead, we regularized the problem by imposing a biologically (developmentally) motivated structure on the covariance matrix, and utilized the information from the paired genotypes from within each line (with and without the *Ubx*[1] mutation). Assuming leg segments are nested within leg-type we consider the line effects of the nine traits to be:

$$\mathbf{G} = \mathbf{B}_{3 \times 3} \otimes \mathbf{W}_{3 \times 3}$$

where **W** captures *within* leg patterns of covariation across leg-segments, and **B** captures *between* leg-type patterns of covariation across thoracic segments. We use subscripts T2, Wt & Ubx to denote the leg-types T2 (Wt), T3(Wt) and T3(*Ubx*).

This model assumes that the covariance structure of the genetic (line) effects for leg-segments are proportional within leg-types. Not all diagonal elements of **B** are identifiable and so we set $b_{T2,T2}$ to one such that **W** is the estimated covariance matrix of line effects within T2 ($\mathbf{G}_{T2}$). The estimated covariance matrix of line effects within T3(Wt) is then $b_{Wt,Wt}\mathbf{W}$, and $b_{Ubx,Ubx}\mathbf{W}$ is the estimated covariance matrix of line effects within T3(*Ubx*). We also assume that the correlation structure of the line effects for leg-segments between leg-types is proportional to the within leg-type correlation structure. Accordingly $b_{T2,Wt}\mathbf{W}$ is the genetic covariance matrix between leg-segments in T2 and T3(Wt), $b_{T2,Ubx}\mathbf{W}$ is the genetic covariance matrix between leg-segments in T2 and T3(*Ubx*) and $b_{Wt,Ubx}\mathbf{W}$ is the genetic covariance matrix between leg-segments in T3(Wt) and T3(*Ubx*).

We entertained two alternate hypotheses for comparison to the unconstrained (but structured) form of **G** described above. First, where the *Ubx*[1] mutation causes a complete transformation of the covariance structure $\mathbf{G}^{(Ubx=1)}$, and secondly a "null" hypothesis where the *Ubx*[1] mutation has no effect on the covariance structure $\mathbf{G}^{(Ubx=0)}$. If the mutation does not influence the covariance structure then we expect the covariance patterns for T3 to be identical for the wild-type and the *Ubx*[1] mutant, and thus constrain parameters in the model such that $b_{Wt,Wt}=b_{Ubx,Ubx}$, $b_{T2,Wt}=b_{T2,Ubx}$ and $b_{Wt,Ubx}=1$. We denote **G** under this set of constraints as $\mathbf{G}^{(Ubx=0)}$. Conversely, if the *Ubx*[1] mutation completely transforms T3 to T2 then we would expect that $b_{T2,T2}=b_{Ubx,Ubx}$, $b_{T2,Wt}=b_{Ubx,Wt}$ and $b_{T2,Ubx}=1$. We denote **G** under this set of constraints as $\mathbf{G}^{(Ubx=1)}$. Thus we compared the model where the components for **G** were estimated as described above, and compared to the constrained models $\mathbf{G}^{(Ubx=1)}$, and $\mathbf{G}^{(Ubx=0)}$. The distribution of the LRT probability values under both hypotheses was relatively uniform, $\mathbf{G}^{(Ubx=0)}$ and $\mathbf{G}^{(Ubx=1)}$, but with slightly inflated type-I errors with the probabilities falling below 0.05 in 7.3% and 5.3% of cases respectively under parametric bootstrap (simulation). In all cases this was close, but significantly greater that the 5% expected.

Environmental Model:

We fitted two classes of model for patterns of environmental covariation; which we refer to as 'structured' and 'full'. In the structured model, the environmental covariances have the same pattern as in the genetic structure:

$$\mathbf{E} = \mathbf{A}_{3x3} \otimes \mathbf{R}_{3x3}$$

where **R** is the within individual matrix of covariances and **A** is the *among* individual matrix of covariances. Because an individual cannot be both Wt and $Ubx^1$, $a_{T2,Ubx}$ and $a_{Wt,Ubx}$ cannot be estimated and were set to zero. We denote the constrained model where $a_{Ubx,Ubx}=a_{Wt,Wt}$ as $\mathbf{E}^{(Ubx=0)}$ and the constrained model where $a_{Ubx,Ubx}=1$ as $\mathbf{E}^{(Ubx=1)}$.

Given the much greater replication at the individual level (>2000) we also entertained a 'full' model in which the within leg-type covariance matrices across leg-types, and the covariance matrix between leg-segments measured on T2 and T3 for wild-type flies, differ from proportionality:

$$\mathbf{E} = \mathbf{W}_{6x6} \oplus \mathbf{U}_{3x3}$$

where **U** is the covariance between the three leg-segments on T3 of $Ubx^1$ flies and **W** is the covariance between the six leg-type by leg-segment combinations in *wild-type* flies. **W** can be represented by the partitioned matrix:

$$\mathbf{W}_{6x6} = \begin{bmatrix} \mathbf{W}_{T2,T2} & \mathbf{W}_{T2,Wt} \\ \mathbf{W}_{Wt,T2} & \mathbf{W}_{Wt,Wt} \end{bmatrix}$$

If the *Ubx¹* mutant has no effect on the residual covariances then we can set $\mathbf{U}=\mathbf{W}_{Wt,Wt}$, to obtain $\mathbf{E}^{(Ubx=0)}$. In contrast if the *Ubx¹* mutant results in a complete homeotic transformation of T3 to T2 then we can set $\mathbf{U}=\mathbf{W}_{T2,T2}$ to obtain $\mathbf{E}^{(Ubx=1)}$.

The two hypotheses, Ubx=0 and Ubx=1, were tested against the unconstrained model using a likelihood ratio test (LRT). Given the large number of parameters and the modest replication at the strain level the asymptotic properties of the LRT may not hold. In order to gauge whether this was an issue we simulated a thousand data-sets under the various constrained models (i.e. Ubx=0 or Ubx=1), fitted a constrained and unconstrained model to each, and then tested for uniformity of the probabilities from the LRT.

### The influence of *Ubx*[1] on T1:

The above analyses do not address the fact that *Ubx*[1] may just be a mutation that effects all legs generally, rather than homeotically transforming one segment (T3) into another (T2). To address this possibility we also report an equivalent analysis for T1 leg-segments in the presence and absence of the *Ubx*[1] mutation.

## *Results*

### The influence of *Ubx¹* on the T3 ectopic apical bristle:

To confirm that the *Ubx*[1] was inducing a homeotic transformation, we examined all individuals for the presence of an ectopic apical bristle on the tibia of the leg on the third thoracic segment, in addition to its normal presence on the second leg. This provided a proxy to determine if the third thoracic segments were "transformed" towards second segmental identities. Consistent with previous observations we found a low overall frequency of this transformation (Rozowski and Akam 2002). However our results demonstrate both segregating genetic variation (strain) and rearing temperature influence the frequency of observing this phenotype (Figure 1). In particular, several of the strains had a high frequency for the presence of the apical bristle, despite the majority showing no expression of this phenotype. This is interesting given that all strains showed varying degrees of the subtle "wing-like" bristles present on the haltere, and an increase in haltere size when measured in $Ubx^1$ individuals (not shown), similar to previous observations (Gibson and van Helden 1997).

These observations were confirmed using a mixed model with a logit link function to assess the fit of models including the effects of temperature. The results of the model suggest that the homeotic transformation, as indicated by the presence of the apical bristle on the third leg, increased in flies reared at 18°C to ~5% (95% CI 3-9%) relative to ~ 1% (0.8-3.4%) reared at 25°C. These values are consistent with previous studies (Rozowski and Akam 2002). In addition there was considerable variation among lines in the penetrance of this phenotype (Figure 1), although with little evidence for interactions between line and temperature (Supplementary Table 1). In no case was an apical bristle observed on wild-type flies ($Ubx^+/Ubx^+$). These results demonstrate that the $Ubx^1$ allele can cause the third-leg to take on second-leg-like developmental fates, which itself is influenced by allelic variation segregating in natural populations.

**The influence of *$Ubx^1$* on the leg segments of T3 in relation to the T2 wild-type:**
Mean Model:

Given the observation that $Ubx^1$ influences the phenotypic expression of the apical bristle, consistent with a homeotic transformation, we also wanted to confirm the effects for quantitative variation on the legs of *Drosophila* (Rozowski and Akam 2002; Stern 2003; Davis et al. 2007), where we were ultimately interested in the change in patterns of phenotypic and genetic co-variation. The results for groups of terms are summarized in the ANOVA table (Tables 1 & 2) and show that the effect of $Ubx^1$ on T3 was strong, particularly for the genotype by leg segment interaction, indicating that it is the relative sizes of the leg-segments that are mainly altered. These results are consistent with the influence of Ubx function in *D. melanogaster* (Stern 2003; Davis et al. 2007), and other species of insects (Mahfooz et al. 2007; Khila et al. 2009). For all measured leg segments on the second and third leg, the $Ubx^1$ mutation influenced the size of each organ in a manner generally consistent with the homeotic transformation (Figure 2). Temperature had a large effect, particularly on the relative sizes of the leg-segments in different legs, although the effect of the $Ubx^1$ the mutation in this context was relatively weak and only marginally significant (Tables 1 & 2).

The results for individual terms are summarized in Table 2. They show that at 18°C the femur and tibia of T3 in the presence of $Ubx^1$ are indistinguishable from the femur and tibia on T2 wild-types, but are very different from those on T3 wild-types. The tarsus of T3 in the presence of $Ubx^1$ was intermediate between T2 and T3 wild-types and significantly different from both. The effect of temperature on T3 in the presence of $Ubx^1$ was more similar to that of T2 wild-types for femur and tibia and indistinguishable from both. However, the tarsus on T3, with or without the $Ubx^1$ mutant, increased in length at higher temperatures where as the tarsus on T2 decreased. Nevertheless, the tarsus length of T3 with the $Ubx^1$ mutant remained intermediate between the T2 and T3 tarsi of wild-types even at 25°C.

As seen in figure 2, all leg segments on T3 are changing their relative lengths consistent with the homeotic influences of the $Ubx^1$ allele. However, the leg segments may be demonstrating differing degrees of transformation. Future studies could consider examining the complex interplay between trait specific genetic background effects, and naturally segregating variation for the relative length of leg segments between T2 and T3.

Reported fixed effects pertain to an unconstrained model for **G** and an unconstrained full model for **E** (see below).

## *Genetic Model: minor influence of $Ubx^1$ on the structure of* **G**.

Here we report the results of the genetic model when **E** was fitted as an unconstrained full model and (separated by a backslash) as an unconstrained structured model. In this model **G** for T3 leg-segments in the presence of the $Ubx^1$ mutation had intermediate scaling between T2 and the T3 wild-type and to a modest degree, covaried more with the T2 leg-segments than were the T3 leg-segments of the wild type (Table 3). In addition, the genetic covariance between L3 leg segments in the presence/absence of the $Ubx^1$ mutation ($b_{Wt,Ubx}$) was closer to $b_{T2,Wt}$ than to one as would also be predicted from a homeotic transformation of **G** (Table 3). However, the differences were subtle with large standard errors, suggesting that an inference of homeotic transformation of the structure of **G** via *Ubx¹* is not necessarily warranted.

We explicitly compared the relative fits of the unconstrained model **G** with a model with parameters constrained assuming that *Ubx¹* had no effect on the genetic covariances $\mathbf{G}^{(Ubx=0)}$ (see methods or table X for details). The difference in fits between G and $\mathbf{G}^{(Ubx=0)}$ was marginal ($\chi_3^2 = 6.72/8.18, p = 0.081/0.042$), with the relevant elements of $\mathbf{G}^{(Ubx=0)}$ being $b_{T2,Wt/Ubx} = 0.792\pm0.068/0.799\pm0.067$ and $b_{Wt/Ubx,Wt/Ubx} = 0.860\pm0.122/0.866\pm0.120$. The discrepancy between the fits of **G** and $\mathbf{G}^{(Ubx=0)}$ seems to be driven by the correlation between Wt and Ubx being less than unity (i.e. $b_{Wt,Ubx}/\sqrt{b_{Wt,Wt}b_{Ubx,Ubx}} \neq 1$) ($\chi_2^2 = 6.05/6.23, p = 0.049/0.044$), since neither constraining the genetic correlation between T2 and Wt and between T2 and Ubx to be equal (i.e. $b_{T2,Ubx}/\sqrt{b_{T2,T2}b_{Ubx,Ubx}} = b_{T2,Wt}/\sqrt{b_{T2,T2}b_{Wt,Wt}}$) nor constraining $b_{Wt,Wt} = b_{Ubx,Ubx}$ produced large changes in the likelihood ($\chi_2^2 = 0.64/1.21, p = 0.725/0.546$ and $\chi_2^2 = 0.51/1.20, p = 0.777/0.550$ respectively). This suggests that while *Ubx¹* is having a subtle influence on the structure of **G**, it is not necessarily consistent with the predicted effects of homeosis.

A global test of **G** versus the developmental model of a complete transformation of T3 to T2, $\mathbf{G}^{(Ubx=1)}$ was more firmly rejected ($\chi_3^2 = 105.86/101.79, p < 0.001$) with the relevant elements of $\mathbf{G}^{(Ubx=1)}$ being $b_{T2/Ubx,Wt} = 0.899\pm0.062/0.885\pm0.058$ and $b_{Wt,Wt} = 0.983\pm0.122/0.929\pm0.111$. While there is suggestive evidence for some increased shared structure in **G** due to *Ubx¹*, the test of **G** vs. $\mathbf{G}^{(Ubx=1)}$ is inconsistent with the mutation causing a complete transformation of the breeding values of T3 into T2.

*Environmental Model:*

In addition to examining how the Ubx1 influences the structure of **G**, we also examined how it influenced covariances for **E**. Here we report the results of the environmental model when **G** was fitted without constraints. In the structured model we found that **E** for T3 leg-segments in the presence of the *Ubx¹* mutation had ~equivalent scaling to the T3 wild-type ($a_{Wt,Wt}$ = 0.991±0.033 and $a_{Ubx,Ubx}$ = 0.991±0.037). The covariance between T2 and T3 leg-segments in the wild-type was moderate ($a_{T2,Wt}$ = 0.416±0.017). There was no evidence that **E** differed from $\mathbf{E}^{(Ubx=0)}$ in a global test ($\chi_1^2 < 0.01, p = 0.994$) with $a_{Ubx/Wt,Ubx/Wt}$= 0.991±0.030 under the $\mathbf{E}^{(Ubx=0)}$ constraint, nor was there evidence that **E** differed from $\mathbf{E}^{(Ubx=1)}$ ($\chi_1^2 = 0.07, p = 0.793$). The large sample sizes and the fact that the greatest difference in log-likelihood between **E**, $\mathbf{E}^{(Ubx=0)}$ and $\mathbf{E}^{(Ubx=1)}$ is 0.034 indicates that the volume of the hyper-ellipsoid for **E** across thoracic segments is virtually identical and therefore is not expected to change considerably under homeotic transformation.

A model in which **E** was fully parameterised provided a significantly better fit to the data than a structured **E** without constraints ($\chi_{48}^2 = 617.32, p < 0.001$) indicating that **E** matrices within or between leg-types do not satisfy the proportionality assumption. In the unconstrained model **E** for the T3 leg-segments of the *Ubx¹* mutant (**U**) was intermediate between $\mathbf{W}_{T2,T2}$ and $\mathbf{W}_{Wt,Wt}$, but generally closer to $\mathbf{W}_{Wt,Wt}$ than $\mathbf{W}_{T2,T2}$: the average of the absolute deviations between elements of **U** and $\mathbf{W}_{T2,T2}$ was 0.187, which although greater than that between **U** and $\mathbf{W}_{Wt,Wt}$ (0.130) was considerably smaller than that between $\mathbf{W}_{T2,T2}$ and $\mathbf{W}_{Wt,Wt}$ (0.308). These differences were small however, and the hypothesis that Ubx=0 (**U**=$\mathbf{W}_{Wt,Wt}$) was not rejected ($\chi_6^2 = 4.35, p = 0.630$) whereas the hypothesis that Ubx=1 was rejected, albeit not strongly ($\chi_6^2 = 13.16, p = 0.041$). Consequently, the large change in likelihood between the structured and full parameterisation of **E** seems to stem from the fact that **E** matrices between leg-types are not proportional to the **E** matrices within the three leg-types which are, broadly speaking, proportional to each other.

The model with the smallest AIC was one in which **E** was unstructured but with the constraint Ubx=0, and **G** had the Ubx=0 constraint except the correlation between T3 Wt and T3 *Ubx¹* leg segments was allowed to be less than unity (0.964±0.018).

### The influence of *Ubx*[1] on T1:

Mean Model:

As predicted by its developmental role, the *Ubx*[1] mutation had negligible (and non-significant) influences on the lengths of leg segments on T1 (Tables 4 & 5). These results are consistent with previously demonstrated effects where reduction of Ubx function had little effect in T1 legs, in *D. melanogaster* (Stern 2003; Davis et al. 2007), as well as several other species of insects (Mahfooz et al. 2007; Khila et al. 2009).

Genetic Model:

For the T1 model, **B** is a 2x2 matrix for the wild-type and the *Ubx*[1] mutant. As before $b_{Wt,Wt}$ was set to one to give $b_{Ubx,Ubx}$=0.825±0.111/0.822±0.110 and $b_{Ubx,Wt}$=0.864±0.063/0.862±0.063 indicating that the genetic (co)variances are reduced in magnitude in the *Ubx*[1] mutant, but with large standard errors on these estimates. While a global test of the null hypothesis Ubx=0 was rejected ($\chi_2^2 = 8.63/8.68, p = 0.013/0.013$), again the discrepancy between **G** and $\mathbf{G}^{(Ubx=0)}$ is driven by the correlation between Wt and Ubx being less than unity (i.e. $b_{Wt,Ubx}/\sqrt{b_{Wt,Wt}b_{Ubx,Ubx}} \neq 1$) ($\chi_1^2 = 5.04/5.05, p = 0.025/0.025$) rather than constraining $b_{Wt,Wt} = b_{Ubx,Ubx}$ ($\chi_1^2 = 2.07/2.15, p = 0.150/0.143$).

*Environmental Model:*

Here we report the results of the environmental model when **G** was fitted without constraints. In the structured model we found that **E** for T1 leg-segments in the presence of the *Ubx*[1] mutation had almost identical scaling to the T1 wild-type ($a_{Ubx,Ubx}$ = 1.011±0.037) and there was no evidence that **E** differed from $\mathbf{E}^{(Ubx=0)}$ ($\chi_1^2 = 0.09, p = 0.759$).

A model in which **E** was fully parameterised provided a significantly better fit to the data than a structured **E** without constraints ($\chi^2_5 = 11.50, p = 0.042$) indicating that **E** for T1 in the two mutants do not satisfy the proportionality assumption and may differ. However, given the large sample sizes the differences were relatively minor and a more direct test of Ubx=0 (i.e. with $a_{Ubx,Ubx}=1$) was less convincing ($\chi^2_6 = 11.60, p = 0.072$).

As with the model of T2 and T3, the T1 model with the smallest AIC was one in which **E** was unstructured but with the constraint Ubx=0, and **G** had the Ubx=0 constraint except the genetic correlation between Wt and *Ubx$^1$* leg segments was allowed to be less than unity (0.946±0.028).

## *Discussion*

Observed patterns of variation in natural and experimental populations have often suggested that there is substantial stasis for many traits. In an evolutionary quantitative genetics framework, such observations have traditionally been explained as either the result of selection (such as the depletion of genetic variation due to historical directional selection, or current stabilizing selection), or as a feature of the genetic architecture of the traits, and in particular with a "genetic constraint" imposed by reduced dimensionality of the **G**-matrix, although evidence for such reductions have been mixed (Mezey and Houle 2005; Mcguigan and Blows 2007; Hansen and Houle 2008; Agrawal and Stinchcombe 2009; Kirkpatrick 2009). Indeed, a number of examples of trait covariation with apparent phenotypic stasis express considerable genetic variation as demonstrated with artificial selection (Weber 1990, 1992; Conner 2002; Frankino et al. 2005; Hansen and Houle 2008; Conner et al. 2011). Yet our understanding of constraint, and changes in patterns of covariation more generally, has been hindered by a lack of *a priori* predictions in most cases.

In this study we developed an experimental system with clear *a priori* predictions for the structure of covariation based on the influence of a homeotic mutation. Despite clear evidence for transformation of the third thoracic leg towards second-leg-like fates (Figure 1 & 2, Tables 1 & 2), we observed weak evidence for changes in the covariance structure for the phenotypic or genetic co-variance matrices (**G** and **E**), counter to these predictions. Thus even with a mutation of profound developmental effect that might be expected to increase covariation among traits on the two segments (since they are developmentally "more similar"), little effect was observed. If such results can be generalized, they suggest that the impact of **M** on **G** may be weak even when mutations have profound developmental effects.

In light of the evidence we consider the most likely explanation for our observations is that the set of traits we examined (femur, tibia and tarsus across thoracic segments) are already highly phenotypically integrated. In this case it may not be surprisingly that $Ubx^1$ was unable to make a highly constrained system detectably more so.

Clearly a mutation such as $Ubx^1$ is unrepresentative of most segregating variation present in natural populations. If such a variant did occur, it would likely be removed rapidly by natural selection. However, the goal of this study was not to mirror allelic effects influencing **G** in natural populations, but to experimentally manipulate **G** based on developmental considerations, with clear *a priori* expectations. Indeed, we consider the approach used here analogous to the numerous studies investigating selection against strong deleterious mutations artificially introduced to high frequencies in natural populations. These mutations were used not so much for the biological plausibility of such deleterious alleles reaching high frequency, but to examine how selection proceeds, and to address additional questions (Agrawal and Stinchcombe 2009; Hollis et al. 2009; Maclellan et al. 2009; Arbuthnott and Rundle 2012). In our system, we utilized $Ubx^1$, to induce homeotic transformations, not because of the expectations of such mutations reaching appreciable frequencies in natural populations, but because it allows for the investigation of fundamental questions in evolutionary biology. Nevertheless, it is clear that there is in fact segregating variation for Ubx function within *D. melanogaster* (Gibson and Hogness 1996; Gibson and van Helden 1997; Gibson et al. 1999; Phinchongsakuldit 2003), and it has potentially played a role in phenotypic changes between species (Stern 1998; Khila et al. 2009). Thus alleles that influence the structure of **G** may indeed fix in natural populations, and it is worth considering their impact on **M**

(Agrawal et al. 2001), and evolvability and patterns of phenotypic integration more generally (Pavlicev et al. 2009).

There are several alternative explanations for our observations. Given the small number of genetically independent lines used for this study, it is possible power was insufficient to assess differences in **G** between $Ubx^1$ and wild-type. However, the experimental results showed a remarkable degree of consistency both for **E** (with a much larger sample size) and **G**. Despite the difficulties in estimating variances and covariances (Hayes and Hill 1981; Meyer and Kirkpatrick 2008), and constructing confidence limits, the lack of differences between $Ubx^1$ and wild-type makes it unlikely that this is simply an issue of power. Indeed one of the most powerful aspects of the approach outlined here is it allows for explicit model fit and comparisons within the context of developmental hypotheses and the experimental manipulation. Given the general prediction from the use of the homeotic mutation, this allowed for specific developmentally motivated structured models to be fit and contrasted. However, future studies utilizing an approach similar to the one described here might attain greater precision by backcrossing a homeotic (or other) mutations into a large natural outbred population (but not lines), and proceeding with a classic breeding experiment (e.g. pedigree/animal model, or nested half-sib design), or by applying artificial selection against the genetic correlation (to determine if the rate of response is impeded in the presence of the homeotic allele).

It is also plausible that homeosis induced in the $Ubx^1/Ubx^+$ heterozygotes is too subtle to have sufficient impact on the covariances for the phenotypes measured. While the phenotypic manifestation of partial homeotic transformation of the haltere was complete in the strains examined, this was not the case for the qualitative measure for the on the meta-thoracic leg (presence of the apical bristle). However, the quantitative estimate for the mean changes suggested almost complete homeotic transformation for the femur and tibia, and a partial transformation for the tarsus. While stronger mutations could be used, the evidence of the current study is inconsistent with an explanation based on mutational severity. When reared at a lower temperature (18°C) the penetrance of the qualitative homeotic trait (apical bristle) increased substantially for the $Ubx^1$ treatment. Yet there was little evidence for an effect of an interaction of mutation and temperature on the structure of **G** and **E**. While the genetic co-variance structure did not change substantially, the influence of $Ubx^1$ on the length of segments was consistent with expectations and previous results. Specifically the meso- and meta-thoracic legs (Table 1, Figure 2), but not the pro-thoracic leg, were influenced by the $Ubx^1$ mutation (Rozowski and Akam 2002; Davis et al. 2007). Exploring how this mutation induces homeotic effects, but with such minimal influences on phenotypic and genetic covariances would be an interesting direction for future research. In particular in the light of much previous work demonstrating increases in both genetic and environmental variances for numerous traits when strains are sensitized with mutations (Dworkin 2005a)

The approach used in this study – utilizing a known genetic perturbation to develop clear *a priori* predictions – complements more traditional breeding designs used in evolutionary quantitative genetics. Indeed this experimental design allowed for a set of explicit developmental models for testing, to address the question at hand. While there has been a trend to both larger and considerably more sophisticated experiments that estimate **G**, it is rarely clear to what the estimated matrix should be compared; i.e. there may not be clear null evolutionary hypotheses about the structure of **G**. Mezey and Houle (2005) provided evidence of a large number of dimensions of available genetic variation for wing shape in *D. melanogaster*. In related studies, Weber (1990) and Conner (2002) demonstrated that there was available genetic variation for composite wing or floral traits, despite little observed phenotypic variation within natural populations, and even across closely related taxa. Such findings are often interpreted as the result of natural selection removing individuals with maladaptive trait combinations. The alternative interpretation is that **G** is of reduced rank, with evidence consistent from a number of traits and systems. Unfortunately in these cases, there was essentially a single estimated **G**, with no clear point of comparison. This is most often because it is unclear what the appropriate "null" hypothesis might be, and so only *ad hoc* comparisons are possible, unless sufficient data about the selective history of the population is available. We argue that an approach integrating some knowledge of developmental genetics, and utilizing tools similar to those used in this study can greatly inform such comparisons, and should be considered as an additional tool to help address evolutionary questions as a complement to more traditional studies.

**Acknowledgements:** We would like to thank David Houle, Ellen Larsen and Greg Gibson for discussions related to this project. Thanks to Will Pitchers, Jeff Conner and Chris Chandler for providing feedback on versions of this manuscript. Funding for this project was from the NSF MCB0922344 to ID.

Worley, A. C., D. Houle, and S. C. Barrett. 2003. Consequences of hierarchical allocation for the evolution of life-history traits. Am Nat 161:153-167.

| Term | Df | F.inc | F.con | Pval |
|---|---|---|---|---|
| (Intercept) | 1 | 49010 | 49010 | <0.001 |
| Thoracic.segment | 1 | 331.9 | 267.4 | <0.001 |
| Genotype | 1 | 4.335 | 12.2 | 0.002 |
| Leg.segment | 2 | 9952 | 9952 | <0.001 |
| Temp | 1 | 0 | 0 | 0.99 |
| Thoracic.segment:Leg.segment | 2 | 173.5 | 244.6 | <0.001 |
| Genotype:Leg.segment | 2 | 194.6 | 194.9 | <0.001 |
| Thoracic.segment:Temp | 1 | 0.065 | 0.24 | 0.63 |
| Genotype:Temp | 1 | 6.415 | 4.97 | 0.035 |
| Leg.segment:Temp | 2 | 512.3 | 512.3 | <0.001 |
| Thoracic.segment:Leg.segment:Temp | 2 | 80.07 | 73.07 | <0.001 |
| Genotype:Leg.segment:Temp | 2 | 2.565 | 2.56 | 0.096 |

Table 1: ANOVA table for the fixed effects of the model examining the influence of the $Ubx^1$ mutation (genotype) and rearing temperature on the length of leg segments within T3 and T2 thoracic segments. F.inc and F.con refer to incremental and conditional F statistics.

| Parameter | Effect | SE | T3_Ubx SE | T3_Ubx Pval |
|---|---|---|---|---|
| Tibia:T3_Ubx | 60.81 | 0.303 | | |
| Tibia:T3_Wt | 61.65 | 0.286 | 0.165 | <0.001 |
| Tibia:T2_Wt | 60.83 | 0.310 | 0.217 | 0.91 |
| Tarsus:T3_Ubx | 28.29 | 0.197 | | |
| Tarsus:T3_Wt | 29.14 | 0.188 | 0.114 | <0.001 |
| Tarsus:T2_Wt | 27.05 | 0.200 | 0.142 | <0.001 |
| Femur:T3_Ubx | 66.52 | 0.289 | | |
| Femur:T3_Wt | 65.61 | 0.273 | 0.151 | <0.001 |
| Femur:T2_Wt | 66.29 | 0.296 | 0.203 | 0.27 |
| Tibia:T3_Ubx:25°C | -1.68 | 0.147 | | |
| Tibia:T3_Wt:25°C | -2.30 | 0.143 | 0.204 | 0.005 |
| Tibia:T2_Wt:25°C | -1.61 | 0.149 | 0.208 | 0.73 |
| Tarsus:T3_Ubx:25°C | 0.92 | 0.101 | | |
| Tarsus:T3_Wt:25°C | 0.64 | 0.104 | 0.144 | 0.06 |
| Tarsus:T2_Wt:25°C | -0.32 | 0.093 | 0.137 | <0.001 |
| Femur:T3_Ubx:25°C | -1.78 | 0.131 | | |
| Femur:T3_Wt:25°C | -2.07 | 0.130 | 0.184 | 0.13 |
| Femur:T2_Wt:25°C | -1.57 | 0.135 | 0.188 | 0.27 |

Table 2: Parameter estimates for the fixed effects of the model examining the influence of the $Ubx^1$ mutation and rearing temperature (18°C vs 25°C) on the length of leg segments within T3 and T2 thoracic segments. Pairs of consecutive non-null cells in columns T3_Ubx SE and T3_Ubx_Pval are the standard error and significance of the effects relative to the preceding effect on T3 in the presence of $Ubx^1$. All p-values were generated from parametric bootstraps.

| parameter | Fit with **E** unconstrained (±SE) | Fit with **E** unconstrained & structured (±SE) | Meaning of term | Hypothesis: complete homeosis $G^{(Ubx=1)}$ | Hypothesis: No homeosis $G^{(Ubx=0)}$ |
|---|---|---|---|---|---|
| $b_{T2,T2}$ | 1 (fixed) | 1 (fixed) | $G_{T2}=W$ | | |
| $b_{Wt,Wt}$ | 0.849±0.125 | 0.842±0.120 | $G_{T3}=b_{Wt,Wt}W$ | | |
| $b_{Ubx,Ubx}$ | 0.915±0.140 | 0.950±0.148 | $G_{Ubx}=b_{Ubx,Ubx}W$ | $=b_{T2,T2}$ | $=b_{Wt,Wt}$ |
| $b_{T2,Wt}$ | 0.780±0.071 | 0.785±0.068 | | | |
| $b_{T2,Ubx}$ | 0.820±0.074 | 0.829±0.076 | | $=1$ | $=b_{T2,Wt}$ |
| $b_{Wt,Ubx}$ | 0.854±0.121 | 0.863±0.121 | | $=b_{T2,Wt}$ | $=1$ |

Table X: Estimated values for the unconstrained, but structured **B** matrix for parameters used in estimating the unconstrained form of **G**. $b_{T2,T2}$ is fixed to one (not estimated) in this structured matrix. $G_{T2}$, $G_{T3}$ and $G_{Ubx}$ are the genetic covariance matrices within leg for the wild-type second, wild-type third and *Ubx¹* third legs respectively. We estimated these parameters with two different **E** matrices, either fitted as an unconstrained full model or as an unconstrained structured model (second and third column). Predicted values under the two alternate hypotheses $G^{(Ubx=1)}$ and $G^{(Ubx=0)}$ represent what parameter values would be estimated to be for **G** if the given hypothesis was true.

| Term | Df | F.inc | F.con | Pval |
|---|---|---|---|---|
| Leg.segment | 3 | 18560 | 18560 | <0.001 |
| Genotype | 1 | 0.361 | 0.356 | 0.56 |
| Temp | 1 | 55.21 | 55.20 | <0.001 |
| Leg.segment:Genotype | 2 | 0.812 | 0.780 | 0.47 |
| Leg.segment:Temp | 2 | 775.2 | 775.2 | <0.001 |
| Leg.segment:Genotype:Temp | 3 | 0.810 | 0.810 | 0.50 |

Table 4: The $Ubx^1$ mutation has little effect on the mean leg segment lengths on the first thoracic segment. ANOVA table for the fixed effects of the model examining the influence of the $Ubx^1$ mutation and rearing temperature on the length of leg segments within T1. F.inc and F.con refer to incremental and conditional F statistics.

| Term | Effect | SE | *Ubx* SE | *Ubx* Pval |
|---|---|---|---|---|
| Tarsus:Genotype_*Ubx* | 17.397 | 0.124 | | |
| Tarsus:Genotype_T1Wt | 17.460 | 0.134 | 0.165 | 0.706 |
| Tibia:Genotype_*Ubx* | 36.201 | 0.221 | | |
| Tibia:Genotype_T1Wt | 36.348 | 0.235 | 0.315 | 0.644 |
| Femur:Genotype_*Ubx* | 53.209 | 0.250 | | |
| Femur:Genotype_T1Wt | 53.338 | 0.275 | 0.122 | 0.302 |
| Tarsus:Genotype_*Ubx*:Temp_25 | 0.494 | 0.077 | | |
| Tarsus:Genotype_T1Wt:Temp_25 | 0.434 | 0.075 | 0.151 | 0.694 |
| Tibia:Genotype_*Ubx*:Temp_25 | 3.842 | 0.172 | | |
| Tibia:Genotype_T1Wt:Temp_25 | 3.900 | 0.168 | 0.331 | 0.862 |
| Femur:Genotype_*Ubx*:Temp_25 | -0.743 | 0.110 | | |
| Femur:Genotype_T1Wt:Temp_25 | -0.960 | 0.119 | 0.124 | 0.089 |

Table 5: Rearing temperature, but not the $Ubx^1$ mutation influences mean leg segment lengths on the first thoracic segment. Parameter estimates for the fixed effects of the model examining the influence of the $Ubx^1$ mutation and rearing temperature on the length of leg segments within T1. Non-null cells in columns *Ubx* SE and *Ubx*_Pval are the standard error and significance of the effects relative to the preceding effect in the presence of $Ubx^1$.

| model | AIC | BIC | logLik |
|---|---|---|---|
| ~ 1 + TEMP + (1+TEMP\|LINE) + (1\|REP) | 120.5 | 136.1 | -54.27 |
| ~ 1 + TEMP + (1+TEMP\|LINE) | 118.5 | 131.5 | -54.27 |
| ~ 1+ TEMP + (1\|LINE) | 114.6 | 122.3 | -54.28 |
| ~ 1+ TEMP | 220.0 | 225.1 | -108.0 |
| ~ 1 | 233.0 | 235.5 | -115.5 |

Supplementary Table 1: Examining the influence of genetic variation and rearing temperature on the penetrance of the apical bristle homeotic phenotype. AIC/BIC and log likelihoods according to glmer (lme4).

|  | First leg | | | Second leg | | | Third leg | | |
|---|---|---|---|---|---|---|---|---|---|
|  | Femur | Tibia | Tarsus | Femur | Tibia | Tarsus | Femur | Tibia | Tarsus |
| Femur | 3.01 | 1.48 | 1.04 | 3.69 | 3.00 | 1.17 | 3.08 | 2.77 | 1.37 |
|  | *1.34-5.37* | *0.25-3.03* | *0.27-2.05* | *1.50-2.05* | *1.0-5.92* | *0.17-2.55* | *1.22-5.88* | *0.92-5.27* | *0.17-2.79* |
| Tibia | 0.64 | 1.78 | 0.77 | 2.14 | 2.62 | 1.04 | 1.64 | 2.03 | 1.24 |
|  | *0.26-0.73* | *0.67-3.23* | *0.15-1.53* | *0.59-4.37* | *0.18-5.03* | *0.31-2.07* | *0.28-3.45* | *0.51-4.0* | *0.28-2.38* |
| Tarsus | 0.70 | 0.68 | 0.73 | 1.30 | 1.13 | 0.80 | 1.19 | 1.07 | 0.88 |
|  | *0.43-0.77* | *0.33-0.74* | *0.30-1.33* | *0.30-2.62* | *0.09-2.47* | *0.28-1.51* | *0.29-2.34* | *0.10-2.21* | *0.29-1.65* |
| Femur | 0.93 | 0.70 | 0.66 | 5.29 | 4.62 | 1.67 | 4.29 | 4.05 | 1.98 |
|  | *0.86-0.93* | *0.47-0.79* | *0.37-0.74* | *2.29-9.42* | *1.69-8.61* | *0.27-3.47* | *1.78-7.71* | *1.45-7.51* | *0.47-4.03* |
| Tibia | 0.74 | 0.85 | 0.57 | 0.86 | 5.41 | 1.75 | 3.61 | 4.43 | 2.15 |
|  | *0.57-0.81* | *0.72-0.89* | *0.11-0.68* | *0.74-0.89* | *2.29-9.9* | *0.4-3.55* | *1.23-6.90* | *1.67-8.17* | *0.61-4.19* |
| Tarsus | 0.60 | 0.69 | 0.84 | 0.65 | 0.67 | 1.26 | 1.51 | 1.68 | 1.25 |
|  | *0.21-0.72* | *0.31-0.75* | *0.71-0.86* | *0.25-0.74* | *0.38-0.74* | *0.50-2.33* | *0.39-3.10* | *0.38-3.32* | *0.45-2.35* |
| Femur | 0.89 | 0.62 | 0.70 | 0.94 | 0.78 | 0.67 | 3.97 | 3.50 | 1.75 |
|  | *0.81-0.90* | *0.26-0.72* | *0.41-0.76* | *0.90-0.94* | *0.62-0.82* | *0.41-0.76* | *1.71-7.08* | *1.26-6.39* | *0.46-3.45* |
| Tibia | 0.76 | 0.72 | 0.59 | 0.84 | 0.90 | 0.71 | 0.83 | 4.45 | 2.02 |
|  | *0.59-0.81* | *0.46-0.79* | *0.13-0.68* | *0.71-0.87* | *0.82-0.93* | *0.40-0.77* | *0.72-0.86* | *1.82-7.89* | *0.64-3.93* |
| Tarsus | 0.63 | 0.73 | 0.81 | 0.68 | 0.73 | 0.88 | 0.70 | 0.76 | 1.60 |
|  | *0.19-0.71* | *0.43-0.78* | *0.68-0.85* | *0.40-0.78* | *0.51-0.79* | *0.81-0.91* | *0.45-0.77* | *0.60-0.83* | *0.62-2.86* |

Supplementary Table 2: **G**-matrix for the leg segments of *D. melanogaster*. Diagonal and upper triangular components represent variances & co-variances, while below the diagonal are correlations. 95% highest posterior density credible intervals for each estimate (*italics*). While we provide this G-matrix for interested researchers, we caution that it was estimated based upon a relatively small number of distinct lines and should be used with utmost caution.

Figure 1: Penetrance of the apical bristle on the third-leg, demonstrating homeotic transformation. A) The morphological marker for the homeotic transformation. The first tibia is from a wild-type second leg (T2), showing the presence of the apical and pre-apical bristles. B) On the tibia of the third leg of wild-type *D. melanogaster* (T3) no apical bristle is present. C) The apical bristle is present on *Ubx$^1$* tibia for the third leg, D) however the penetrance of this phenotype varies across strains.

Figure 2: The degree of homeosis for the meta-thoracic femur, tibia and basi-tarsus due to *Ubx$^1$*. This measure represents the scaled effect of the *Ubx$^1$* mutation for each segment on the meta-thoracic (T3) leg, scaled by the difference between the T2 and T3 segments. While each segment demonstrates evidence of homeotic transformation, they vary, with a proximal to distal decrease in degree of homeosis. Error bars represent 95% CIs for this measure.

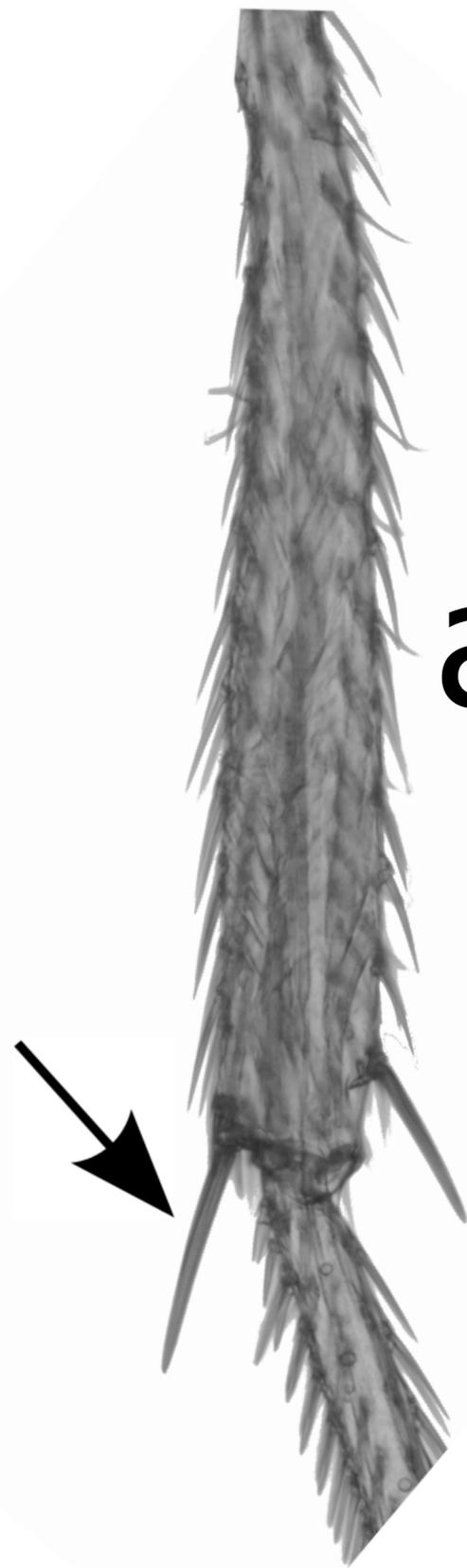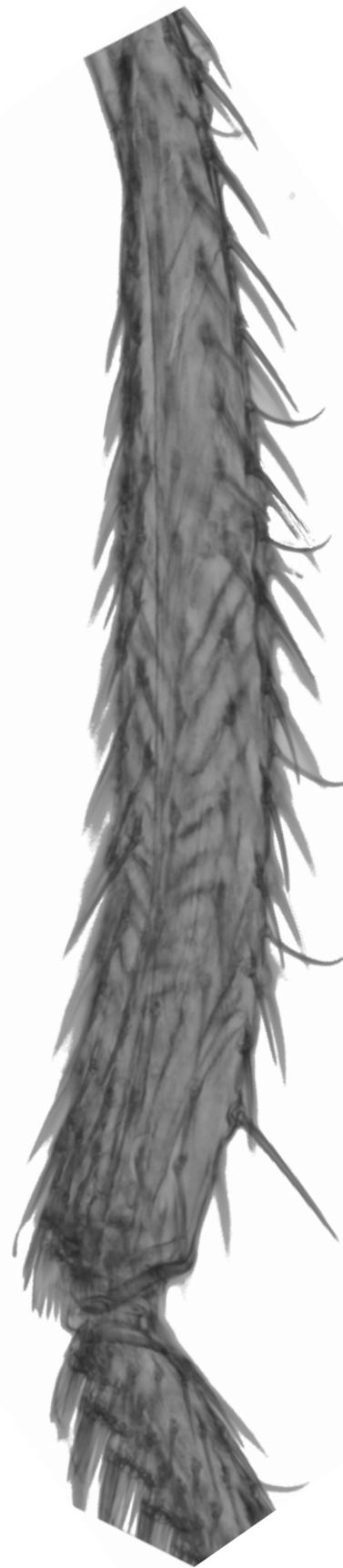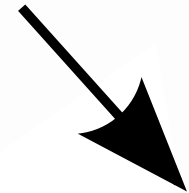

Figure 1a-c

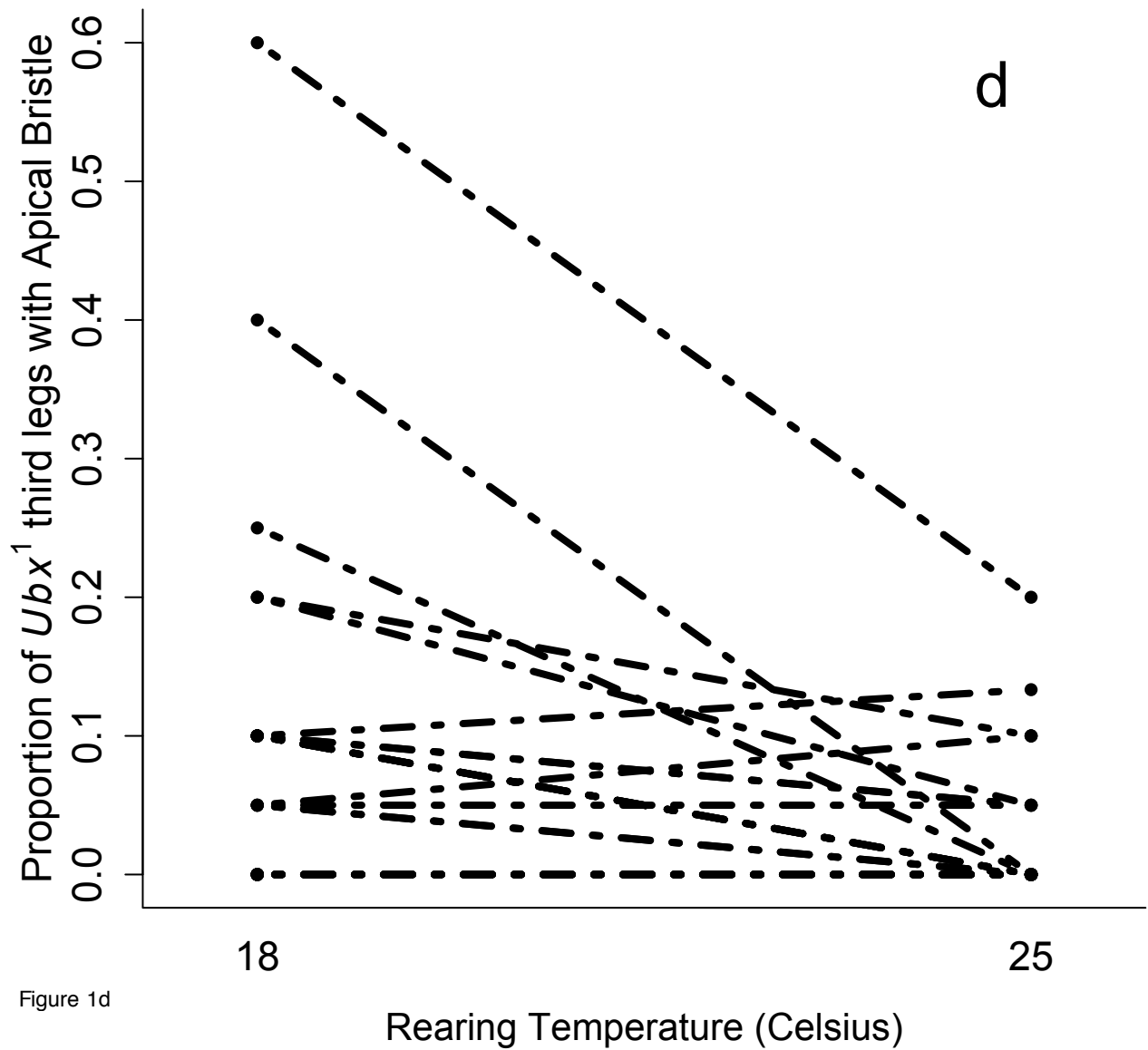

Figure 1d

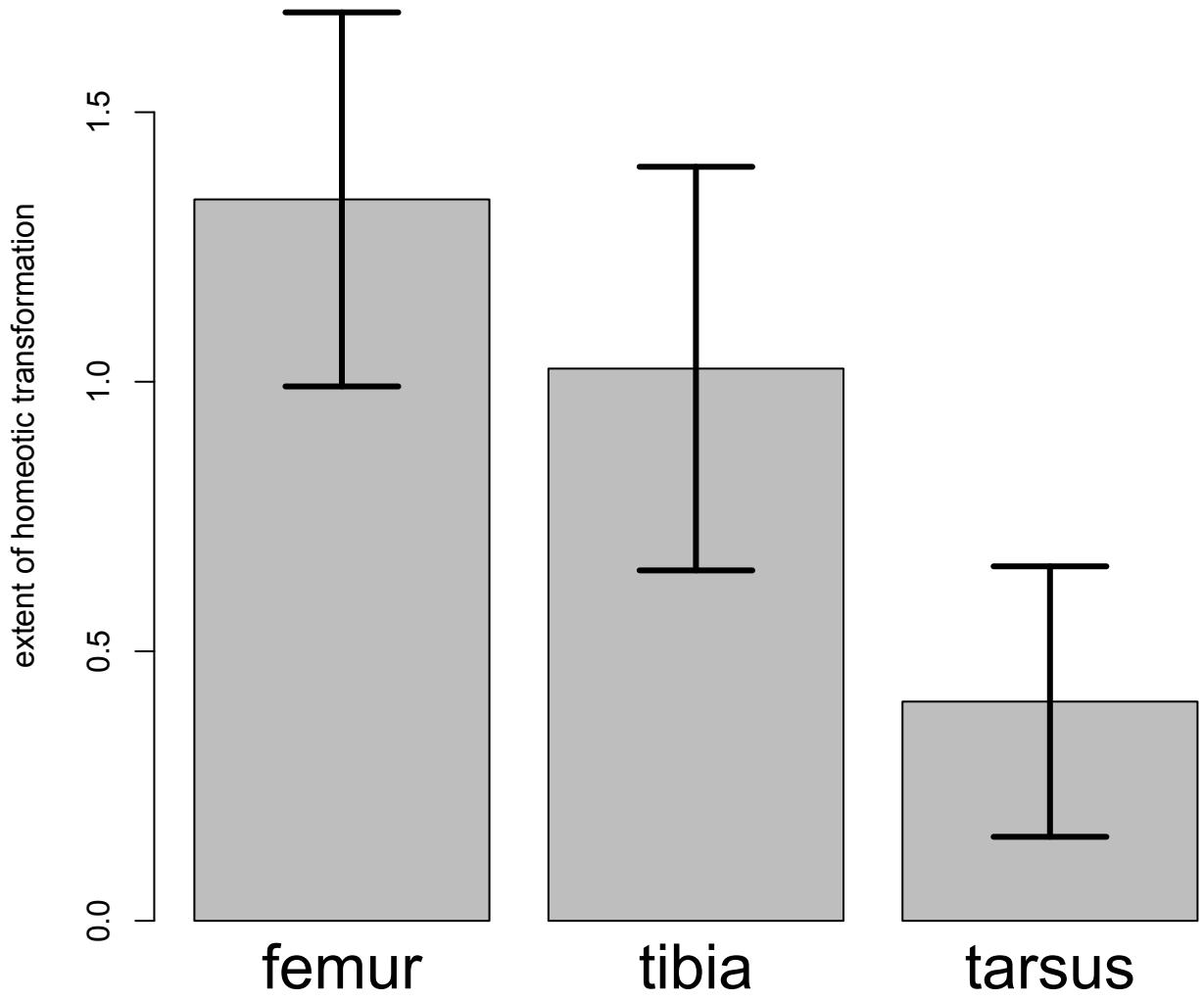

Figure 2